\documentclass[12pt]{revtex4}
\usepackage{epsfig}

\newcommand{\be}{\begin{equation}}
\newcommand{\ee}{\end{equation}}
\newcommand{\bea}{\begin{eqnarray}}
\newcommand{\eea}{\end{eqnarray}}

\begin{document}

\title{\textbf{PATH INTEGRAL REPRESENTATION FOR \\ SPIN SYSTEMS}}

\author{Naoum Karchev }

\affiliation{ Department of Physics, University of Sofia, 1126 Sofia, Bulgaria }

\begin{abstract}
The present paper is a short review of different path integral representations of the partition function of quantum
spin systems. To begin with, I consider coherent states for $SU(2)$ algebra. Different parameterizations of the coherent states lead to different path integral representations. They all are unified within an $U(1)$ gauge theory of quantum spin systems.
\end{abstract}


\maketitle

\section{Introduction}

Path integrals over fields is widely used approach to quantum field theory. I shall discuss the quantum field theory
of spin systems.

Spin systems obtain their magnetic properties from a system of localized
magnetic moments. The dynamical degrees of freedom are spin-s
operators $\textbf{S}_r$ of localized spins which satisfy the $SU(2)$ algebra. It was shown by Haldane \cite{Haldane1,Haldane2}
that quantum spin systems could be formulated in terms of path integrals over vectors which identify
the local orientation of the spin of the localized electrons.

In the present paper I discuss different path integral representations of the partition function of quantum spin systems.

\section{\textbf{Coherent states for $SU(2)$ algebra }}

Let $\text{S}_r^{\alpha}$ are the spin operators with $\alpha=1,2,3$ and $r$ labels
the lattice sites. They obey the $SU(2)$ algebra
\begin{equation}\label{PIR1}
\left[\text{S}_r^{\alpha}\,,\,\text{S}_{r'}^{\beta}\right]\,=\,i\delta_{r,r'}
\epsilon^{\alpha\beta\gamma}S_r^{\gamma}
\end{equation}
We consider the Haisenberg model with a Hamiltonian
\begin{equation}
H\,=\,\sum\limits_{r,r'}J_{r,r'}\textbf{S}_r\cdot \textbf{S}_{r'}
%+\kappa\sum\limits_r \left(\hat S^3_r\right)^2
\end{equation}
where the only condition is $J_{r,r}=0$.

Let us put in correspondence to any lattice site a $(2s+1)$ dimensional
space of $SU(2)$ group representation. Then the Hilbert space of the
system is the tensor product of all these spaces.

Let us choose as a "ground state" $\vert 0\rangle $ the vector which
satisfies
\begin{equation}
\text{S}_r^3\vert 0\rangle\,=\,s\vert 0\rangle
\end{equation}
Following Radcliffe\cite{Radcliffe} one defines the coherent states and the
conjugated coherent states by the relations
\begin{equation}\label{R1}
\vert z\rangle\,=\,e^{\sum\limits_r z_r \text{S}^-_r}\vert 0\rangle\,=
\prod\limits_r e^{z_r \text{S}^-_r}\vert 0\rangle
\end{equation}
\begin{equation}\label{R2}
\langle \bar z \vert\,=\,\langle 0\vert
e^{\sum\limits_r S^+_r\bar z_r}\,=\,
\langle 0\vert\prod\limits_r e^{ S^+_r\bar z_r}
\end{equation}
where $ S_r^{\pm}\,=\, S_r^1\pm i S_r^2$, $z_r$ are complex
numbers and $\bar z_r$ are their complex conjugated.

The follow formulae for the matrix elements are straightforeward
generalization of those given in the Radcliffe's paper \cite{Radcliffe}
\begin{equation}\label{R3}
\langle \bar z'\vert z\rangle\,=\,
\prod\limits_r\left(1+\bar z_r' z_r\right)^{2s}
\end{equation}
\begin{equation}\label{R4}
\langle \bar z'\vert S^-_r\vert z\rangle\,=\,\frac {2s\bar z_r'}
{\left(1+\bar z_r'z_r\right)}\langle \bar z'\vert z\rangle
\end{equation}
\begin{equation}\label{R5}
\langle \bar z'\vert S^+_r\vert z\rangle\,=\,\frac {2s z_r}
{\left(1+\bar z_r'z_r\right)}\langle \bar z'\vert z\rangle
\end{equation}
\begin{equation}\label{R6}
\langle \bar z'\vert S^3_r\vert z\rangle\,=\,
\frac {s\left(1-\bar z_r' z_r\right)}
{\left(1+\bar z_r'z_r\right)}\langle \bar z'\vert z\rangle
\end{equation}

The "resolution of unity", which is an expression of the identity operator
in terms of the coherent state operators $\vert z\rangle\langle\bar z\vert$
is given by
\begin{equation}\label{R7}
\int\prod\limits_r d\mu(z_r)
\frac {1}{\prod\limits_r\left(1+\bar z_r z_r\right)^{2s}}
\vert z \rangle\langle \bar z \vert\,=1
\end{equation}
where
\begin{equation}\label{R8}
d\mu(z_r)\,=\,\frac {(2s+1)}{\left(1+\bar z_r z_r\right)^2}
\frac {d^2z_r}{\pi}
\end{equation}
and the product $\prod\limits_r$ is over the all lattice sites.

Setting in equations (\ref{R4}),(\ref{R5}) and (\ref{R6}) $z'=z$ one obtains the diagonal matrix
elements of the generators $\bf{S}_r$
\begin{equation}\label{R9}
\langle \bar z\vert \textbf{S}_r\vert z\rangle\,=\,s\textbf{n}_r
\langle \bar z\vert z\rangle
\end{equation}
where $\bf{n}_r$ are unit vectors $(\textbf{n}_r^2=1)$ given by
\begin{equation}\label{R10}
n^1_r=\frac {z_r+\bar z_r}{1+\bar z_r z_r},\quad
n^2_r=\frac 1i \frac {z_r-\bar z_r}{1+\bar z_r z_r},\quad
n^3_r=\frac {1-\bar z_r z_r}{1+\bar z_r z_r},\quad
\end{equation}

The equations (\ref{R10}) map the complex plane onto unit sphere $S^2$. It is
convenient to use  the azimuthal angle
$0\leq\theta_r\leq\pi$ and the polar angle $0\leq\varphi_r\leq2\pi$ as variables determining the coherent states.  Making use of the
stereographic projection
\begin{equation}
z_r=tg \frac {\theta_r}{2}\, e^{i\varphi_r}
\end{equation}
one obtains
$$
\textbf{n}_r=(\cos \varphi_r\sin\theta_r,\,\sin\varphi_r\sin \theta_r,\,
\cos\theta_r)
$$
Now the equations (\ref{R4}-\ref{R7}) can be rewritten in terms of the two angles. For
example the matrix elements (\ref{R3}) take the form
\begin{equation}
\langle\textbf{n}'\vert\textbf{n}\rangle\,=\,\prod\limits_r
e^{i\gamma(\textbf{n}_r',\textbf{n}_r)}
\left(\frac {1+\textbf{n}_r'\cdot\textbf{n}_r}{2}\right)^s
\end{equation}
where $\gamma(\textbf{n}_r',\textbf{n}_r)$ is the area of the spherical triangle
with vertices $\textbf{n}^0=(0,0,1)$, $\textbf{n}_r'$ and $\textbf{n}_r$. The measure
(\ref{R8}) is manifestly rotationally invariant if we rewrite it in terms
of unit vectors
\begin{equation}\label{vec}
d\mu(\textbf{n}_r)\,=\,\frac {2s+1}{4\pi}\sin \theta_r d\theta_r d\varphi_r\,=\,
\frac {2s+1}{4\pi}\delta(\textbf{n}^2_r-1)d^3n_r
\end{equation}
where $\delta$ is Dirac's delta function.

Finally, mapping the complex plane onto disk with radius $\sqrt {2s}$\,\,
one introduces another parametrization of the coherent states
\begin{equation}\label{HP1}
a_r=\frac {\sqrt {2s}\, z_r}{(1+\bar z_r z_r)^{\frac 12}} \qquad
\bar a_r=\frac {\sqrt {2s}\,\bar z_r}{(1+\bar z_r z_r)^{\frac 12}}
\end{equation}
where $a_r$ and $\bar a_r$ are complex numbers subject to the
condition $\bar a_r a_r\leq 2s$. Then the spin vectors $\textbf{S}_r=s\textbf{n}_r\,
\,(\textbf{S}_r^2=s^2)$ are given by
\bea\label{HP2}
S_r^- & = & \bar a_r\sqrt {2s-\bar a_r a_r} \nonumber \\
S_r^+ & = & \sqrt {2s-\bar a_r a_r}\,\, a_r      \\
S_r^3 & = & s-\bar a_r a_r \nonumber
\eea
where $S_r^{\pm}\,=\,S_r^1\,\pm\,iS_r^2$, and the measure (\ref{R8}) takes
the form
\begin{equation}\label{HP3}
d\mu(\bar a_r a_r)\,=\,\frac {2s+1}{2s}\frac {d\bar a_r d a_r}{2\pi i}
\end{equation}

\section{\textbf{Path integral approach for Heisenberg model}}
Following the path integral approach I use the coherent states in the
evaluation of the partition function
\begin{equation}\label{PI1}
{\cal Z}(\beta)\,=\,Tre^{-\beta H}.
\end{equation}
 In equation (\ref{PI1}) $\beta$ is the inverse temperature. It is evident from equation (\ref{R7}) that
this function  admits the representation
\begin{equation}\label{PI2}
{\cal Z}(\beta)\,=\,
\int\prod\limits_r d\mu(z_r)
\frac {1}{\prod\limits_r\left(1+\bar z_r z_r\right)^{2s}}
\langle \bar z \vert e^{-\beta H} \vert z \rangle
\end{equation}

One may consider the operator $e^{-\beta H}$ as a multiple of many
small evolutions
\begin{equation}\label{PI3}
e^{-\beta H}\,=\,\lim_{N\to\infty}\left(1-\frac {\beta}{N} H\right)^N
\end{equation}
Then, using the equation (\ref{R7}) one obtains
\bea\label{PI4}
Tre^{-\beta H}
& = & \lim_{N\to\infty}\int\prod\limits_{r}d\mu(z_r)
\prod\limits_{k=1}^{N-1}d\mu(z_r(\tau_k))
\langle\bar z\vert\left(1-\frac {\beta}{N} H\right)
\vert z(\tau_{N-1})\rangle \nonumber \\
& & \langle \bar z(\tau_{N-1})\vert\left(1-\frac {\beta}{N} H\right)\vert
z(\tau_{N-2})\rangle \dots
\langle \bar z(\tau_1)\vert\left(1-\frac {\beta}{N} H\right)
\vert z\rangle \nonumber \\
& & exp\left\{-2s\sum\limits_r\left[\ln\left(1+\bar z_r z_r\right)
+\ln\left(1+\bar z_r(\tau_{N-1}) z_r(\tau_{N-1})\right)\right.\right. \nonumber\\
& & \left.\left.+\dots+
\ln\left(1+\bar z_r(\tau_1) z_r(\tau_1)\right)\right]\right\}
\eea

The kernel
$\langle \bar z(\tau_k)\vert\left(1-\frac {\beta}{N} H\right)
\vert z(\tau_l)\rangle$
can be represented in the form
\bea\label{PI5}
\langle \bar z(\tau_k)\vert\left(1-\frac {\beta}{N} H\right)
\vert z(\tau_l) \rangle\ & = & \left(1-\frac {\beta}{N}h\left(\bar z(\tau_k),
z(\tau_l)\right)\right)\langle \bar z(\tau_k)\vert z(\tau_l)\rangle \\
 & \simeq & exp\left\{-\frac {\beta}{N} h\left(\bar z(\tau_k),z(\tau_l)
\right)+2s\sum\limits_r\ln\left(1+\bar z_r z_r\right)\right\}\nonumber
\eea
Making  use of the equations (\ref{R3}-\ref{R6}), one represents the Hamiltonian in the form
\bea
& & h\left(\bar z(\tau_k),z(\tau_l)\right)= \\
& & s^2\sum\limits_{r,r'}J_{r,r'}
\frac {2\left[\bar z_r(\tau_k) z_{r'}(\tau_l)
+\bar z_{r'}(\tau_k) z_{r}(\tau_l)\right]+
[1-\bar z_r(\tau_k) z_r(\tau_l)][1-\bar z_{r'}(\tau_k) z_{r'}(\tau_l)]}
{[1+\bar z_r(\tau_k) z_r(\tau_l)][1+\bar z_{r'}
(\tau_k) z_{r'}(\tau_l)]}\nonumber
%-\kappa 2s(2s-1)\sum\limits_r\frac {\bar z_r(\tau_k) z_r(\tau_l)}
%{\left(1+\bar z_r(\tau_k) z_r(\tau_l)\right)^2}.
\eea
where the term independent of $\bar z_r(\tau_k)$ and $z_r(\tau_l)$ is
dropped.

Now we proceed taking the continuum limit $N\to\infty$ and find the
path integral representation of the partition function
\begin{equation}\label{Keller}
{\cal Z}(\beta)\,=\,\int \prod\limits_{\tau,r}d\mu(z_r(\tau))e^{-S(\bar z,z)}
\end{equation}
where
\begin{equation}\label{Keller1}
S(\bar z,z)\,=\,\int\limits^{\beta}_0 d\tau\left\{2s\sum\limits_r
\frac {1}{1+\bar z_r(\tau)z_r(\tau)}\bar z_r\dot z_r+
h(\bar z(\tau),z(\tau))\right\}
\end{equation}
is the action, and the hamiltonian is
\begin{equation}
h(\bar z(\tau),z(\tau))\,=\,s^2\sum\limits_{r,r'} J_{r,r'}
\textbf{n}_r(\tau)\cdot\textbf{n}_{r'}(\tau)
\end{equation}
In the above, the overdots correspond to time derivatives.
The complex fields $\bar z_r(\tau),
z_r(\tau)$, and the real vector fields
$\textbf{n}_r(\tau)$ satisfy periodic boundary conditions
$\bar z_r(\beta)=\bar z_r(0), z_r(\beta)=z_r(0), \textbf{n}_r(0)=\textbf{n}_r(\beta)$.

One can use the coherent states labeled by the unit vector $\textbf{n}_r$ to
derive the path integral representation for the partition function \cite{Haldane1,Haldane2}. Then
the measure is given by
\begin{equation}\label{vec2}
d\mu(\textbf{n})\,=\,\prod\limits_{\tau,r}\frac {2s+1}{4\pi}d^3 n_r(\tau)
\delta(\textbf{n}_r^2(\tau)-1)
\end{equation}
(see Eq.(\ref{vec})) and the action adopts the form
\begin{equation}\label{vec3}
S=\int\limits_0^{\beta} d\tau \left[is\sum\limits_{r}\textbf{A}(\textbf{n}_r)\cdot
{\dot \textbf{n}}_r(\tau)\,+\,h(\tau)\right]
\end{equation}
In equation (\ref{vec3}) $\textbf{A}(\textbf{n}_r)$ is the vector potential of a Dirac magnetic monopole
at the center of the unit sphere
\begin{equation}
\textbf{A}\,=\,\frac {1-\cos\theta}{\sin\theta}\textbf{e}_{\varphi}
\end{equation}
It obeys locally
\begin{equation}
\partial_{\textbf{n}}\times \textbf{A}(\textbf{n})\,=\,\textbf{n}
\end{equation}

The kinetic term in Eq.(\ref{vec3}) is invariant under the gauge transformations
$$
\textbf{A}\to\textbf{A}\,+\,\partial_{\textbf{n}}\alpha,
$$
where the parameter $\alpha$ is defined on the sphere. It is more convenient
for further calculations to do a gauge transformation which leads to the vector potential
\begin{equation}\label{vec4}
\textbf{A}\,=\,-\coth\theta\,\, \textbf {e}_{\varphi}
\end{equation}
In this case the half of the string is up the north pole and the other
half is down the south pole. Thus, the vector potential is an even
function of its argument
\begin{equation}
\textbf {A}(\textbf {n})\,=\,\textbf{A}(-\textbf{n})
\end{equation}

Making use of the third parametrization of the coherent states (\ref{HP1}) one
obtains a path integral in terms of complex fields $a_r(\tau)$ and
$\bar a_r(\tau)$, which satisfy $\bar a_r(\tau)a_r(\tau)\leq2s$ and periodic boundary conditions. The measure
is given by
\begin{equation}\label{HP}
d\mu(\bar a,a)\,=\,\prod\limits_{\tau,r}\frac {2s+1}{2s}
\frac {d \bar a_r(\tau) d a_r(\tau)}{2\pi i}
\end{equation}
Substituting in the Hamiltonian and the kinetic term one rewrites the action in the form
\begin{equation}\label{HPaction}
S=\int\limits_0^{\beta}d\tau\left [\frac 12\sum\limits_r
\left(\bar a_r(\tau)\dot a_r(\tau)-\dot {\bar a}_r(\tau)a_r(\tau)\right)
+\sum\limits_{r,r'}J_{r,r'}\textbf{S}_r(\tau)\cdot\textbf{S}_{r'}(\tau)\right]
\end{equation}
where the spin vectors are given by (2.18).

The field theories defined by the actions (\ref{Keller1}), (\ref{vec3}) and (\ref{HPaction}) can be thought
of as a particular case of a more general Abelian gauge field theory. To
see how does this come about I introduce two complex fields
$\psi_r^k(\tau)$ ($k=1,2$) by the relations
\bea\label{Sb}
|\psi_r^1(\tau)| & = &
\frac {\sqrt {2s}}{(1+\bar z_r(\tau) z_r(\tau))^{\frac 12}} =
(2s-\bar a_r(\tau)a_r(\tau))^{\frac 12} \nonumber \\
\psi_r^2(\tau) & = &
\frac {\sqrt {2s}\,z_r(\tau)}{(1+\bar z_r(\tau) z_r(\tau))^{\frac 12}}
= a_r(\tau)
\eea
Let substitute Eq.(\ref{Sb}) in Eq.(\ref{Keller1}) or Eq.(\ref{HPaction}) and take into
account the condition
\begin{equation}\label{Sbgauge}
\arg\, \psi_r^1(\tau)\,=\,0
\end{equation}
One gets
\begin{equation}\label{Sbaction}
S\,=\,\int\limits_0^{\beta}d\tau\left [\sum\limits_{\alpha,r}\bar \psi_r^{\alpha}(\tau)
\frac {d}{d\tau}\psi_r^{\alpha}(\tau)\,+\,\sum\limits_{r,r'}J_{r,r'}
\textbf{S}_r(\tau)\cdot\textbf{S}_{r'}(\tau)\right]
\end{equation}
where the spin vectors are equal to
\begin{equation}\label{Sbspin}
S^{\nu}_r(\tau)\,=\,\frac 12 \bar \psi_r^{\alpha}(\tau)\ \sigma^{\nu}_{\alpha\alpha'}
\psi_r^{\alpha'}(\tau)
\end{equation}
and $\sigma^{\nu}$ are Pauli matrices.
The new fields obey the constraint \cite{ArAu}
\begin{equation}\label{Sbcon}
\bar \psi_r^1(\tau)\psi_r^1(\tau)\,+\,\bar \psi_r^2(\tau)\psi_r^2(\tau)\,=\,2s .
\end{equation}

Let take care of the local constraint (\ref{Sbcon}) by introducing an extra term
in the action with the Lagrangian multiplier field $\lambda_r(\tau)$ which
enforces the constraint. Collecting all terms one obtains the final
expression for the partition function
\begin{equation}\label{SbZ}
{\cal Z}(\beta)\,=\,\int\prod\limits_{\alpha,r,\tau}d\bar \psi_r^{\alpha}(\tau)d
\psi_r^{\alpha}(\tau)\prod\limits_{r,\tau}\delta\left(arg\psi_r^1(\tau)\right)
e^{-S_{tot}}
\end{equation}
where the action
\bea\label{SbS}
S_{tot}\,  =
\,\int\limits_0^{\beta}d\tau\left [\sum\limits_{\alpha,r}\bar \psi_r^{\alpha}(\tau)
\left (\frac {d}{d\tau}-i\lambda_r(\tau)\right)\psi_r^{\alpha}(\tau)\,+
\,2si\lambda_r(\tau)\right. \nonumber \\
\left. +  \sum\limits_{r,r'}J_{r,r'}
\textbf{S}_r(\tau)\cdot\textbf{S}_{r'}(\tau)\right]
\eea
is invariant under the gauge transformations
\bea\label{SbGt}
\psi_r^{\alpha\prime}(\tau) & =  & e^{i\gamma_r(\tau)}\psi_r^{\alpha}(\tau),\quad
\bar\psi_r^{\alpha\prime}(\tau)\,=\,e^{-i\gamma_r(\tau)}\bar\psi_r^{\alpha}(\tau),\nonumber \\
\lambda_r^{\prime}(\tau) & = & \lambda_r(\tau)\,+\,\frac {d}{d\tau}\gamma_r(\tau)
\eea
if the gauge parameters $\gamma_r(\tau)$ satisfy
$\gamma_r(0)=\gamma_r(\beta)$.

Let us discuss the Abelian gauge theory (\ref{SbS}). Following the standard
procedure of quantization, one has to impose an additional gauge fixing
condition. If we do this, imposing the condition (\ref{Sbgauge}), and then solve the
constraint (\ref{Sbcon}), using different parameters we get different field
theoretical realizations of the Heisenberg model. On the other hand, one
can choose the gauge fixing condition in an alternative way. For example,
a convenient gauge condition is the temporal condition imposed on the
Lagrangian multiplier $\lambda_r(\tau)$. The gauge fixing condition reads
\begin{equation}
\lambda_r(\tau)\,=\,\lambda_r
\end{equation}
where $\lambda_r$ depends on the lattice sites but is not a function of the
imaginary time. It follows from derivation, that
so obtained field-theoretical descriptions are equivalent.

\section{Summary}

Starting from different parameterizations of coherent states of $SU(2)$ algebra I have derived different path integral representations for the quantum spin systems. In the first case Eqs.(\ref{Keller},\ref{Keller1}) the path integral is over complex fields and the representation is appropriate for Monte Carlo numerical calculations. In the second case Eqs. (\ref{vec2},\ref{vec3}), the path integral is over unite vectors which identify
the local orientation of the spin of the localized electrons. For antiferromagnetic systems it is utilized to derive \cite{Haldane1} an effective $\sigma$ model of the antiferromagnetism.
For these calculations one uses the representation Eq.(\ref{vec4}) of the Dirac's vector field. Finally, the path integral over the complex fields
$\bar a_r(\tau)$ and $a_r(\tau)$ Eqs.(\ref{HP},\ref{HPaction}) is an alternative to the operator Holstein-Primakoff approach.

\end{document}